\documentclass[aip, apl, amsmath, amssymb, reprint] {revtex4-1}
\usepackage{graphicx}
\usepackage{dcolumn}
\usepackage{bm}
\usepackage{textgreek}

\usepackage{gensymb}

\usepackage[colorlinks=true, linkcolor=blue, allcolors=blue]{hyperref}%

\newcommand {\pcc}{PCC}
\newcommand {\phc}{PhC}

\begin{document}

\preprint{AIP/123-QED}
\title[Sample title]{Silicon photonic crystal cavities at near band-edge wavelengths}

\author{Salahuddin Nur}
\email{salahuddin.nur.13@ucl.ac.uk}
\altaffiliation[Current address: ]{Institute of Electronics, Atomic Energy Research Establishment, Savar, Dhaka 1349, Bangladesh}
\affiliation{London Centre for Nanotechnology, University College London, London WC1H 0AH, UK}  
\affiliation{Dept.\ of Electronic \& Electrical Engineering, University College London, London WC1E 7JE, UK}

\author{Hee-Jin Lim}%
 \altaffiliation[Current address: ]{Korea Research Institute of Standards and Science, Dajeon 34113, Korea}
\affiliation{London Centre for Nanotechnology, University College London, London WC1H 0AH, UK}

\author{Jeroen Elzerman}
\affiliation{London Centre for Nanotechnology, University College London, London WC1H 0AH, UK}
\affiliation{Dept.\ of Electronic \& Electrical Engineering, University College London, London WC1E 7JE, UK}
\author{John J. L. Morton }
\affiliation{London Centre for Nanotechnology, University College London, London WC1H 0AH, UK}
\affiliation{Dept.\ of Electronic \& Electrical Engineering, University College London, London WC1E 7JE, UK}

\date{\today}

\begin{abstract}

	
We demonstrate photonic crystal L3 cavities with resonant wavelength around 1.078 \textmu m on undoped silicon-on-insulator, designed to enhance spontaneous emission from phosphorus donor-bound excitons.\enspace We have optimised a fabrication recipe using readily available process materials such as polymethyl methacrylate (PMMA) as a soft electron-beam mask and a Chemical Vapour Deposition (CVD) grown oxide layer as a hard mask. Our bilayer resist technique efficiently produces photonic crystal cavities with a quality factor ($Q$) of $\sim 5,000$ at a wavelength of $1.078$ \textmu m, measured using cavity reflection measurements at room temperature. We observe a decrease of $Q$ as the cavity resonance shifts to shorter wavelengths ($Q \lesssim3,000$ at wavelengths $< 1.070$ \textmu m), which is mostly due to the intrinsic absorption of silicon.

\end{abstract}


\maketitle

%



Defect spins in solid state materials are attractive candidates for scalable implementation and integration of quantum information processing (QIP)\cite{Ladd:2010kq, Gordon:2013hs}, metrology \cite{Bonato:2016eu, Wolfowicz:2018ts, Tarasenko:2018ky} and communication systems \cite{Blok:2015di, Johnson:2017jl}. For example, coherent spins in diamond and their interactions with photons have been exploited for optically-mediated entanglement of matter-based systems \cite{Bernien:2013kj, Pfaff:2014hy}. 
While nitrogen-vacancy (NV) centres in diamond possess many attractive features that have underpinned key quantum information/communication demonstrations, some of the optical properties are sub-optimal (broad phonon sideband and spectral broadening) while thin-film growth and fabrication processes still need to be perfected. For such reasons, other materials systems combining excellent optical and spin memory properties with mature fabrication techniques are being explored to develop effective spin-photon interfaces\cite{Gordon:2013hs, Atature:2018hh, Christle:2017cy, Zhong:2015iy}. Amongst these have been vacancies in silicon carbide (SiC) and defects in silicon (Si).
%
Silicon and SiC host defects and impurities with long spin coherence time \cite{Steger:2012ev, Christle:2017cy} and narrow linewidth emission of photons \cite{Yang:2006gz, Thewalt:2007bq, Steger:2011fh, Christle:2017cy} and permit coherent optical control of spins \cite{Zwier:2015go}. These features, combined with the mature industrial techniques in manufacturing and on-chip integration, make such spins attractive for efficient multi-qubit coupling and realising large scale QIP systems. However, strong non-radiative processes in silicon-based host materials restrict fluorescence efficiency \cite{Schmid:1977bi, Thewalt:2007bq, Sumikura:2011ch} and indistinguishable single photon generation \cite{Pelton:2002hv, Christle:2017cy}, thus limiting the potential of optical interfaces with most defects in silicon. This issue can, in principle, be addressed by engineering the local photonic environment in the host material: for example, incorporating photonic structures such as circular Bragg resonators (CBRs) \cite{Davanco:2011iv} or photonic crystal cavities (\pcc s) \cite{Joannopoulos:1997ez, Faraon:2012ky} can enhance photon emission and collection efficiency by several orders of magnitude, potentially allowing it to compete with non-radiative processes such as Auger recombination.

Enhanced light-matter interaction in \pcc s\cite{Joannopoulos:1997ez} has been demonstrated for various quantum emitters including NV centers in diamond \cite{Faraon:2012ky}, rare-earth-doped crystals \cite{Zhong:2015iy} and quantum dots in GaAs\cite{Reinhard:2011er}, with observed improvements in the radiative emission\cite{Zhong:2015iy, Lee:2015ig}. Similar schemes with \pcc s can be utilised to enhance defect related emission in Si \cite{Fu:2004kp} and SiC \cite{Calusine:2014gv, Calusine:2016hr} systems. Several defects such as shallow donors in silicon \cite{Dean:1967dl, Yang:2006gz, Thewalt:2007bq} and divacancy\cite{Christle:2017cy, Lohrmann:2017kz}, transition metal  \cite{Baur:1997cx, Son:1999en}, and Ky5 color centre \cite{Son:1996cg, Calusine:2016hr} defects in SiC \cite{Lohrmann:2017kz} manifest spin-coupled optical emission with wavelength near $1.078$ \textmu m (i.e.\ near the silicon band-edge). \pcc s with a high ratio of Q-factor to mode volume ($Q/V_{m}$) could be used to develop efficient spin-photon interfaces to such defects, however, experimental studies in this wavelength range on such materials are limited. In SiC, planar \pcc s with wavelengths in the range 1100--1300~nm have been fabricated\cite{Calusine:2014gv, Calusine:2016hr} with $Q/V_{m} \sim 900 - 1500 (n/\lambda)^3$, but the band-gap of Si is substantially smaller and there are significant challenges related to absorption when the \pcc\  resonance approaches the band-gap energy.
Photon absorption from the host material leads to uncertainties in the photonic modes and photonic band-gap as compared to non-dispersive materials\cite{Sakoda:2005fr}, and detection of weak optical signals near the silicon band-edge is made more challenging due to the low quantum efficiency of Si detectors or the high dark count in InGaAs detectors\cite{Dussault:2004fu}. 

Here, we design, fabricate and study suspended Si \pcc s, made from silicon-on-insulator (SOI) substrates, with wavelengths in the range 1065--1085 nm at room temperature. We incorporate fine-tuning and band-folding in our L3 cavities in order to achieve high quality factors and better extraction of light, which could be utilised to enhance shallow impurity spontaneous emissions such as donor bound exciton ($^{31}$P  D$^{0}$X $\rightarrow$ D$^{0}$) transitions in silicon. We have optimised a fabrication recipe using relatively inexpensive process materials in order to realise Si \pcc s using a conventional etching/pattern transfer processes. Finally, we used cross-polarisation confocal microscopy\cite{Faraon:2008da, Galli:2009ge} with a broadband source and a spectrograph with low-noise Si detector array to measure \pcc\ reflection spectra, observing \pcc\  Q-factors in the range 2,600 to 6,200, increasing with cavity wavelength.



\begin{figure}[t]
	\centerline{\includegraphics[width=8.6cm]{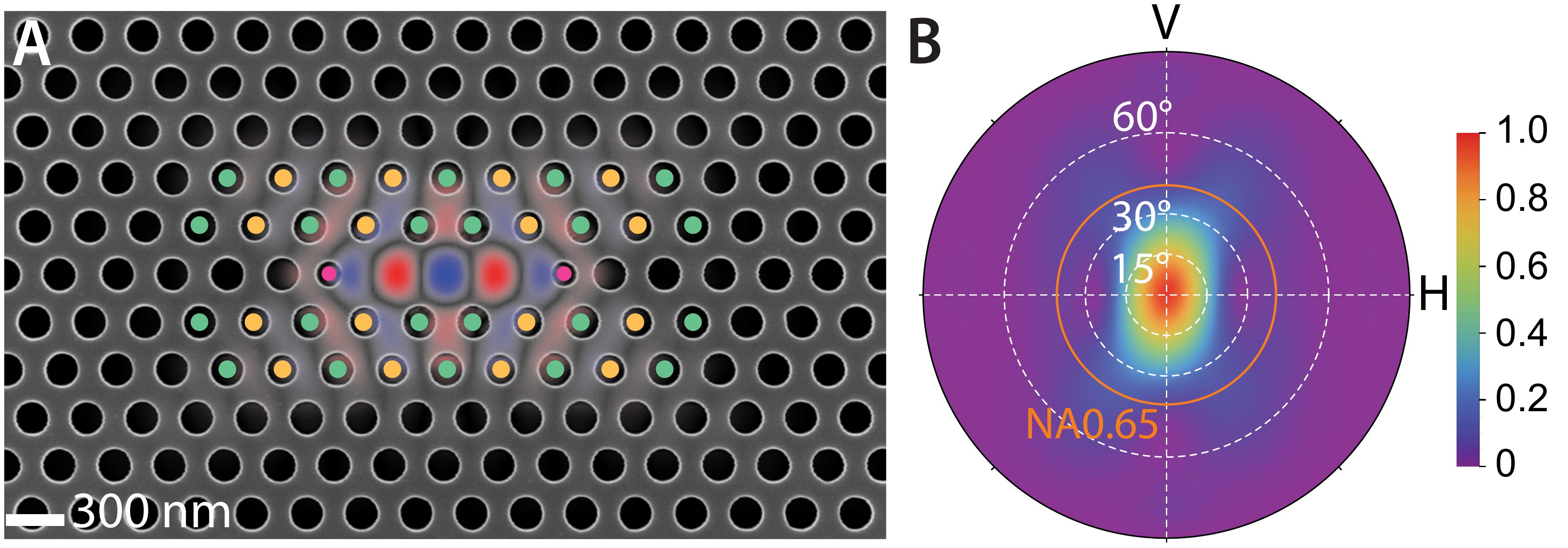}}
 	\caption{ 
	\textbf{(a)} Scanning electron microscope (SEM) image of a typical fabricated L3 cavity, with superimposed simulated electric field profile based on the contour data extracted from the SEM image. Green and yellow circles indicate holes whose radii are adjusted to improve collection efficiency (see main text). 
	\textbf{(b)} Far field profile\cite{Demarest:1996cu} of the fundamental mode of the L3 cavity calculated by the contour FDTD method\cite{Taflove:2004wc}.
	}
	\label{fig1:fab_far}
 \end{figure}

L3 cavities are implemented by removing a row of three air holes from the hexagonal photonic crystal (\phc) lattice with lattice constant $a$.\enspace We systematically study near Si band-edge resonant modes by  fabricating L3 \pcc s with lattice constants ranging from $a$ = 240 to 300 nm. The L3 fundamental mode ($L3^0$)$^[$\cite{Chalcraft:2007hu}$^]$ can be further engineered to improve $Q$ and light outcoupling, while keeping the mode volume below $ 0.9 (\lambda/n)^3$. $Q$ can be increased by changing position and/or size of one or more side-holes adjacent to the cavity\cite{Akahane:2005ge}. The position displacement of any side hole from its original location in the lattice is indicated by a shift, $\Delta s_i$ and any absolute change in the corresponding hole radius by, $\Delta r_{\textrm{side},i}$. Figure~\ref{fig1:fab_far}(a) shows the SEM image of an L3 cavity where the position and size of a pair of holes (marked in red) on either side of the cavity has been adjusted ($\Delta s_{\textrm{1}} = 0.16a$ and $\Delta r_{\textrm{side,1}} = +0.06a$). 
Such a change in the design can produce $Q$ as high as $\sim 45,000$ $^[$\cite{Akahane:2005ge}$^]$.\enspace 

The vertical collection efficiency ($\eta$) is improved by implementing a band folding scheme\cite{Tran:2009vd, Portalupi:2010tr} in which gratings of periodicity $2a$ are superimposed on the \phc\ lattice by modulating the radii of certain holes in the vicinity of the cavity. In Figure~\ref{fig1:fab_far}(a), the radius of the green set of holes above and below the cavity is increased ($\Delta r = +0.02 a$) from the regular air hole radius of $r_{0} = 0.28a$ while the radius of the yellow marked holes is reduced ($\Delta r = -0.02 a$). The incorporation of such a hole-size modulation in the design limits $Q \sim 1.5\times10^4$, but increases $\eta$ up to $\sim$0.8 for an NA = 0.65 (see Figure~\ref{fig1:fab_far}(b)), where NA is the numerical aperture of the collection objective. These mode properties were simulated based on data extracted from SEM images of fabricated structures, and are similar to those based on the idealised design. This indicates fabrication errors are unlikely to be a significant contribution to collection losses. We have also implemented designs with three side hole shifts ($\Delta s_{\textrm{1}} = 0.17a$, $\Delta s_{\textrm{2}} = -0.025a$ and $\Delta s_{\textrm{3}} = 0.17a$) and a slightly modified modulation scheme in which the radii of the green holes remain at $r_{0}$, while the yellow have a larger radius ($\Delta r = +0.02 a$). This modified hole modulation scheme, in principle, can further improve $Q$ while maintaining collection efficiencies of $\eta \sim 0.8$ for an NA = 0.65.



\begin{figure}[t]
	\centerline{\includegraphics[scale=1]{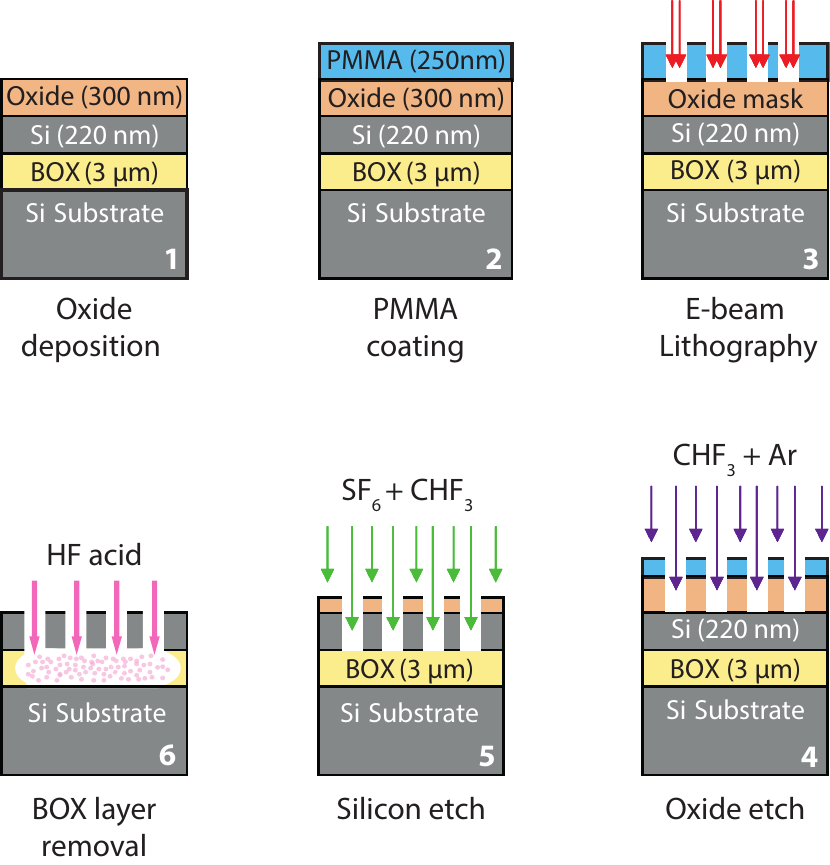}}
 	\caption{ 
	Summary of the process steps for fabricating Si \pcc s with the biilayer (PMMA/Oxide (PECVD)) resist (see text for full details).
	}
	\label{fig2:fab_process}
 \end{figure}

We  optimised fabrication process steps to transfer e-beam lithography (EBL, at 30 kV) profiles into the thin (220 nm) Si device layer of the SOI chip with minimum distortions. Effective realisation of Si \phc s with small lattice constants ($< 300$ nm) depends on the availability of a lithographic mask that can withstand plasma etching long enough to transfer the patterns efficiently to the 220~nm Si layer \cite{BaileyIII:1998da}. We adopted a bilayer resist of polymethyl methacrylate (250 nm) and PECVD grown oxide layer (300~nm), which is less affected by proximity effects and can provide sufficient etch selectivity and anisotropy for the plasma etch steps under consideration.
The process recipe includes two steps of reactive ion etching (RIE): CHF$_{3}$/Ar plasma to transfer the pattern on the oxide layer and CHF$_{3}$/SF$_{6}$ plasma to etch the silicon layer. Conditions for anisotropic etching have been obtained by further adjusting RIE parameters including flow rates (CHF$_{3}$ - 25~sccm \& Ar - 25 sccm), RF power (150W), and pressure (30~mT) for oxide etch and flow rates (CHF$_{3}$ - 58~sccm \& SF$_{6}$  - 25~sccm), RF power (150W), and pressure (10~mT) for silicon etch. We run a cooling step (50~sccm Ar flow without plasma) for two minutes after each 30 s long plasma etching step to avoid PMMA deformations by heat, and repeat this cycle 24 times until the pattern penetrates through the 300~nm oxide hard mask layer. Finally, the pattern is transferred to the 220 nm Si device layer using plasma etch along with the oxide hard mask.\enspace To release the suspended membrane containing the \phc, we undercut the buried oxide (BOX) layer of the SOI chip and remove remaining oxide masks together with hydrofluoric acid (HF). The major process steps in the optimised fabrication recipe are shown in Figure~\ref{fig2:fab_process}. When the fabricated devices are inspected under SEM, it is found that fabrication errors are small. The hole radii in the silicon membrane were larger than the intended values by less than 10~nm, producing good \phc s with lattice constants $a$  between 240 and 300~nm.


\begin{figure}[t]
	\centerline{\includegraphics[width=8.5cm]{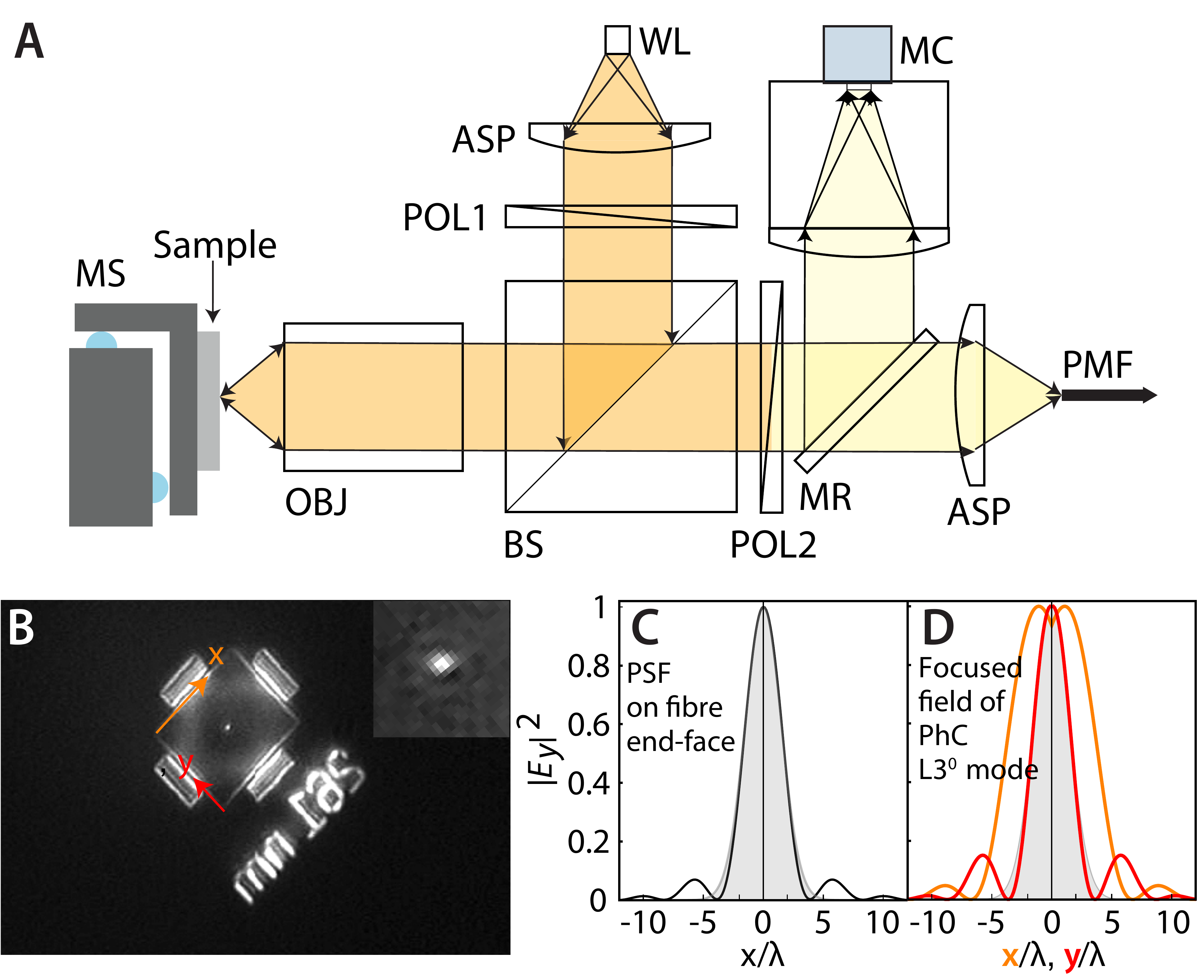}}
	\caption{
	\textbf{(a)}
	Confocal microscopy setup with cross-polarisation to measure the \pcc\ reflection spectrum. \textbf{(b)} The inset below shows an optical image of a \pcc\ (appearing as a bright dot in centre of the rectangle) captured by a high sensitivity, monochrome camera (MC). The MC was incorporated for focussing/sample surface inspection and placed after an analyser (POL2) which is aligned orthogonal to the polariser (POL1).
	Definitions used are MS: motorised stage, 
	OBJ: objective lens, 
	BS: non-polarising beam splitter, 
	ASP: aspheric lens, 
	WL: white light source, MR: flippable mirror, and PMF: polarisation-maintaining fibre.
	\textbf{(c)} The point spread function (PSF) at the PMF end-face. 
	\textbf{(d)} The focused field intensity based on the far-field of cavity mode (see Fig.~\ref{fig1:fab_far}(b)). The orange and red curves denote the focused field intensity distribution along $x$ and $y$-direction, respectively. The Grey shaded curves in (c, d) are mode profiles of fibre mode.
	}
	\label{fig3:setup}
\end{figure}


Fabricated Si \pcc s were characterised by cavity reflection measurements using a cross-polarisation confocal setup shown in Figure~\ref{fig3:setup}(a). An optical image of a PCC (a
bright dot in centre of the rectangle) captured by the CMOS camera (CM) is shown in Figure~\ref{fig3:setup}(b). The cavity reflection signal was collected into a single mode polarisation maintaining fibre (PMF) in a confocal configuration where the point spread function is matched to the fibre mode, as shown in Figure~\ref{fig3:setup}(c), with a spatial resolution \cite{Novotny:2006ex}, $R_{\textrm{spatial}} \approx 0.6\lambda_{\mathrm{exc}}/\mathrm{NA} \approx 1$~\textmu m determined for a high numerical aperture (NA = 0.65) and an excitation wavelength ($\lambda_{\mathrm{exc}}$) of $\sim 1.078$ \textmu m. The maximum coupling efficiency was measured to be 70\% using the Gaussian beam collimated from a single mode fibre\cite{KowaleviczJr:2006wc}.
The cross-polarisation setup was implemented by setting the two polarisers, POL1 and POL2 in orthogonal directions \cite{Faraon:2008da, Galli:2009ge} (Figure~\ref{fig3:setup}(a)). A suppression ratio around $10^6$ has been measured with this system, which in turn allows to select reflections associated with the cavity mode. The bright spot shown in the camera image (Figure~\ref{fig3:setup}(b)) also contains signals from higher order modes and tails of the cavity resonance, resolved with a signal-to-noise ratio (SNR) of $\sim35$. In Figure~\ref{fig3:setup}(c), the black trace is the point spread function/illumination profile from the source observed at the polarisation-maintaining fibre (PMF) end, the grey area denotes the collection by the PMF. The orange and red traces in Figure~\ref{fig3:setup}(d) are the focussed field intensities of $L3^{0}$ mode along $x$ and $y$-axes, respectively. The $L3^{0}$ field profile at the PMF end is calculated from field amplitudes and phases of the simulated far-field profile \cite{Novotny:2006ex} of the cavity (Figure~\ref{fig1:fab_far}(b)). Now, cavity scattering makes the focused spot spread wider along $x$-axis (orange trace) than along $y$-axis (red trace) at the end of the collection PMF. This gives a broader shape for the $L3^{0}$ field profile along $x$-axis (orange trace in Figure~\ref{fig3:setup}(d)) than the collection by the PMF (grey shaded curves in Figure~\ref{fig3:setup}(c) \& (d)) and causes a mode mismatch with the PMF. From the traces in Figure~\ref{fig3:setup}(d), we extract a mode mismatch of $\sim 30\%$. Finally, by taking account of the collection efficiency of \pcc s ($\eta \sim 80\%$), path losses in the optical setup, PMF coupling efficiency etc., a total coupling efficiency of $\sim 9\%$ has been estimated for our confocal setup.

An Acton SP-2750 Princeton Instruments spectrometer with a focal length of $0.75 \;$m was used to capture the spectrum of the collected signal. We used a 300~g/mm blaze grating optimised for wavelengths around 1~\textmu m which can provide a resolution of approximately $150 \;$pm around the $^{31}$P  D$^{0}$X $\rightarrow$ D$^{0}$ transition wavelengths. The dispersed light from the grating was detected by a Si-based CCD array (PyLoN System Silicon CCD Camera) enabling the measurement of quality factors up to $\sim 10,000$. The output power of the LED source used (M1050D1, Thorlabs) peaks at 50~mW, but varies considerably across the measurement window of interest (1060--1090~nm), falling to  about 10~mW at the $^{31}$P  D$^{0}$X $\rightarrow$ D$^{0}$ transition wavelength.




\begin{figure}[t]
	\centerline{\includegraphics[width=8.5cm]{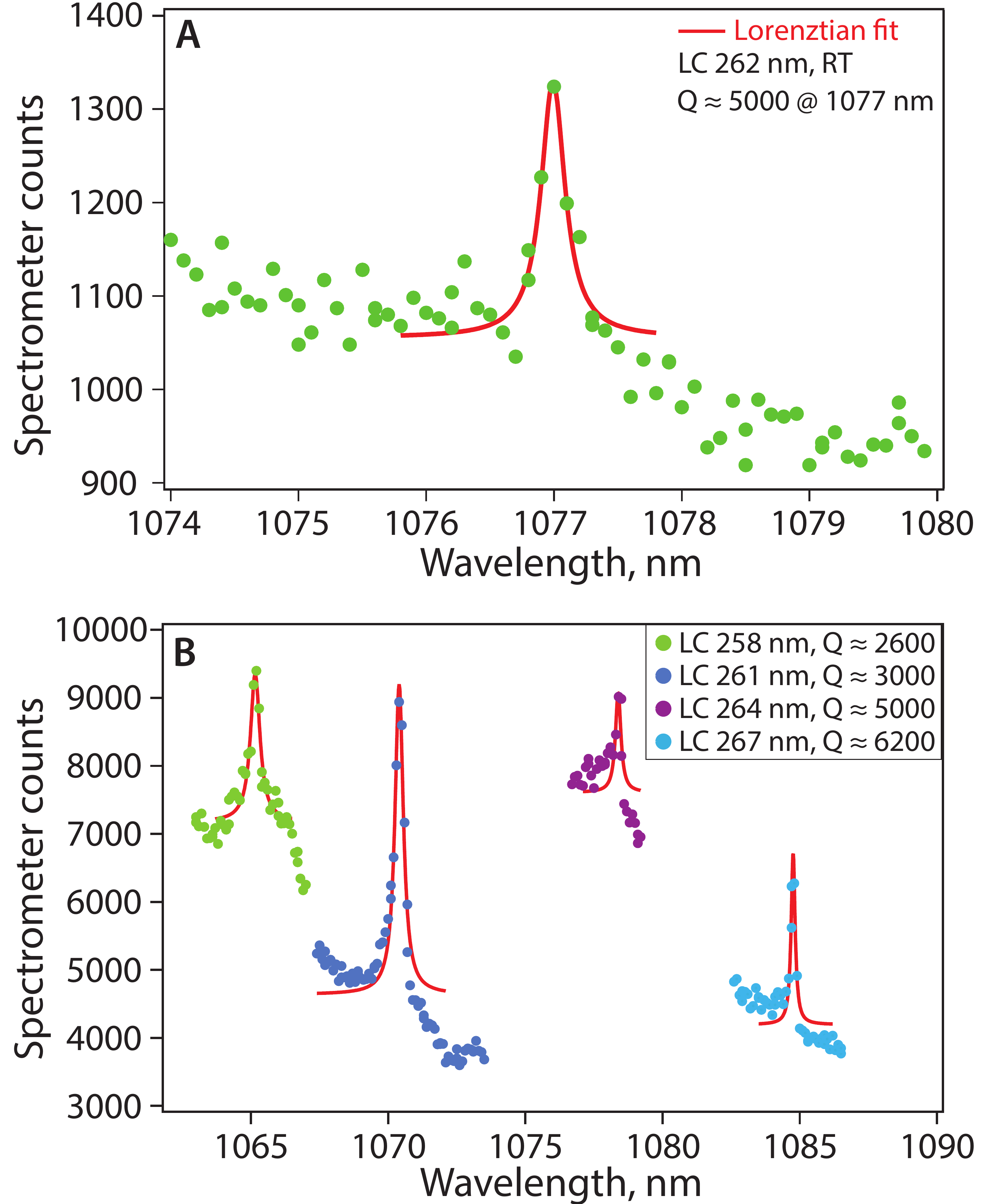}}
	\caption{
	Room temperature cavity reflection spectra measured with a cross-polarisation setup. Lorentzian fits to $L3^0$ modes reveal quality factors for L3 \pcc s tuned with (a) single side-hole shift ($\Delta s_{\textrm{1}} = 0.16a$ and $\Delta r_{\textrm{side1}} = 0.06a$), (b) three side-hole shifts ($\Delta s_{\textrm{1}} = 0.17a$, $\Delta s_{\textrm{2}} = -0.025a$ and $\Delta s_{\textrm{3}} = 0.17a$) for a collection efficiency of $\eta \sim 0.8$. 
	}
	\label{fig4:RS_measurement}
\end{figure}


Results of cavity reflection measurements at room temperature are summarised in Figure~\ref{fig4:RS_measurement}. For a single side-hole shifted L3 cavity with $a = 262$ nm, we observe the fundamental resonant mode ($L3^0$) appearing at a wavelength of $\sim 1077$ nm, which closely matches the D$^{0}$X transitions in silicon. The designed $Q$ for this $L3^0$ mode is $Q_{\mathrm{des}}\approx 29,000$, while the Lorentzian fit to the measured data gives an experimental $Q_{\mathrm{ex}}\approx5,000$. Such a mismatch is often attributed to fabrication or structural imperfections in the \pcc \cite{Maeno:2017dv}, however, for near band-edge Si \pcc s, the intrinsic material absorption can also be a dominant loss mechanism and this was not accounted for in the simulations of the idealised structures.
Silicon has an absorption coefficient $\gtrsim 2 \;$cm$^{-1}$) near the band-edge, at room temperature\cite{Komma:2016vb, Ross:2017tv}, giving an upper bound of only few thousand for achievable \pcc ~$Q$ values. To investigate this in greater depth, we studied three side-hole shifted L3 cavities, designed for $Q_{\mathrm{des}} \approx 50,000$, across a lithographic tuning range with a step size of 3 nm. 

Near Si band-edge resonant modes for the three side-hole shifted L3 cavities are shown in  Figure~\ref{fig4:RS_measurement}(b), with selected fundamental resonances ($L3^0$) at $1065$ nm, $1070.4 \; $nm, $1078.4 \; $nm and $1084.8 \; $nm for L3 cavities with lattice constants of $258$ nm, $261 \; $nm, $264 \; $nm and $267 \; $nm, respectively. 
The quality factors ($Q_{\mathrm{ex}}$) extracted from measurement data are $\sim 2,600$, $\sim 3,000$, $\sim 5,000$ and $\sim 6,200$, growing with increasing $L3^0$ wavelength. 
At room temperature, optical absorption in Si drops gradually with decreasing photon energies below the bandgap \cite{Komma:2016vb, Ross:2017tv}, consistent with our measurements. We also note that  both Si \pcc ~designs (single- and three- side-hold shifts) show $Q_{\mathrm{ex}}\approx 5,000$ for $L3^0$ near $^{31}$P D$^{0}$X transition wavelengths ($\sim 1078$ nm), suggesting that losses in our near band-edge cavities at room temperature are dominated by intrinsic material absorption in Si. Absorption at these wavelengths in Si drops significantly at cryogenic temperatures \cite{Komma:2016vb, Ross:2017tv}, and so significantly higher Q values can be expected. Low temperature PL measurements on $^{31}$P doped float zone (Fz) type Si sample, using a similar measurement setup, have shown the detection of approximately 1000 effective $^{31}$P D$^{0}$X emitters in Si (ensemble linewidth of $\sim0.1\;$nm). Even the Q-factors of around 5000 that we measure at room temperature should be sufficient for detecting optical emission from spins down to the single or few donor/defect level in Si\cite{Dean:1967dl, Yang:2006gz} or SiC \cite{Lohrmann:2017kz}.

In summary, we have optimised a low cost fabrication process recipe for realising Si \pcc s and efficiently fabricated L3 cavities with resonances near silicon band-edge wavelengths. Room temperature optical characterisation of fabricated cavities unveils an absorption-limited $Q$ of 5000 with a collection efficiency close to $80\%$ around 1078~nm. Such near band-edge Si \pcc s may play an important role in realising efficient spin-photon interfaces in Si and SiC systems.


\begin{acknowledgments}
We thank the London Centre for Nanotechnology cleanroom staff for their technical support. We thank G. Matmon, UCL, for use of experimental equipment and insightful comments, and T. F. Krauss and C. Reardon, University of York for providing the SOI wafer and for useful discussions. The research leading to these results has received funding from the European Research Council under the European Union's Seventh Framework Programme (FP7/2007-2013)/ ERC grant agreement No. 279781 (ASCENT) and Horizon 2020 research and innovation programme (grant agreement No. 771493 (LOQO-MOTIONS). We also acknowledge financial support from EPSRC and UCL Engineering. 
\end{acknowledgments}


\bibliography{bibliography_28Sep2018}

\begin{thebibliography}{49}%
\makeatletter
\providecommand \@ifxundefined [1]{%
 \@ifx{#1\undefined}
}%
\providecommand \@ifnum [1]{%
 \ifnum #1\expandafter \@firstoftwo
 \else \expandafter \@secondoftwo
 \fi
}%
\providecommand \@ifx [1]{%
 \ifx #1\expandafter \@firstoftwo
 \else \expandafter \@secondoftwo
 \fi
}%
\providecommand \natexlab [1]{#1}%
\providecommand \enquote  [1]{``#1''}%
\providecommand \bibnamefont  [1]{#1}%
\providecommand \bibfnamefont [1]{#1}%
\providecommand \citenamefont [1]{#1}%
\providecommand \href@noop [0]{\@secondoftwo}%
\providecommand \href [0]{\begingroup \@sanitize@url \@href}%
\providecommand \@href[1]{\@@startlink{#1}\@@href}%
\providecommand \@@href[1]{\endgroup#1\@@endlink}%
\providecommand \@sanitize@url [0]{\catcode `\\12\catcode `\$12\catcode
  `\&12\catcode `\#12\catcode `\^12\catcode `\_12\catcode `\%12\relax}%
\providecommand \@@startlink[1]{}%
\providecommand \@@endlink[0]{}%
\providecommand \url  [0]{\begingroup\@sanitize@url \@url }%
\providecommand \@url [1]{\endgroup\@href {#1}{\urlprefix }}%
\providecommand \urlprefix  [0]{URL }%
\providecommand \Eprint [0]{\href }%
\providecommand \doibase [0]{http://dx.doi.org/}%
\providecommand \selectlanguage [0]{\@gobble}%
\providecommand \bibinfo  [0]{\@secondoftwo}%
\providecommand \bibfield  [0]{\@secondoftwo}%
\providecommand \translation [1]{[#1]}%
\providecommand \BibitemOpen [0]{}%
\providecommand \bibitemStop [0]{}%
\providecommand \bibitemNoStop [0]{.\EOS\space}%
\providecommand \EOS [0]{\spacefactor3000\relax}%
\providecommand \BibitemShut  [1]{\csname bibitem#1\endcsname}%
\let\auto@bib@innerbib\@empty
\bibitem [{\citenamefont {Ladd}\ \emph {et~al.}(2010)\citenamefont {Ladd},
  \citenamefont {Jelezko}, \citenamefont {Laflamme}, \citenamefont {Nakamura},
  \citenamefont {Monroe},\ and\ \citenamefont
  {O{\textquoteright}Brien}}]{Ladd:2010kq}%
  \BibitemOpen
  \bibfield  {author} {\bibinfo {author} {\bibfnamefont {T.~D.}\ \bibnamefont
  {Ladd}}, \bibinfo {author} {\bibfnamefont {F.}~\bibnamefont {Jelezko}},
  \bibinfo {author} {\bibfnamefont {R.}~\bibnamefont {Laflamme}}, \bibinfo
  {author} {\bibfnamefont {Y.}~\bibnamefont {Nakamura}}, \bibinfo {author}
  {\bibfnamefont {C.}~\bibnamefont {Monroe}}, \ and\ \bibinfo {author}
  {\bibfnamefont {J.~L.}\ \bibnamefont {O{\textquoteright}Brien}},\ }\href
  {\doibase 10.1038/nature08812} {\bibfield  {journal} {\bibinfo  {journal}
  {Nature}\ }\textbf {\bibinfo {volume} {464}},\ \bibinfo {pages} {45}
  (\bibinfo {year} {2010})}\BibitemShut {NoStop}%
\bibitem [{\citenamefont {Gordon}\ \emph {et~al.}(2013)\citenamefont {Gordon},
  \citenamefont {Weber}, \citenamefont {Varley}, \citenamefont {Janotti},
  \citenamefont {Awschalom},\ and\ \citenamefont {Van~de
  Walle}}]{Gordon:2013hs}%
  \BibitemOpen
  \bibfield  {author} {\bibinfo {author} {\bibfnamefont {L.}~\bibnamefont
  {Gordon}}, \bibinfo {author} {\bibfnamefont {J.~R.}\ \bibnamefont {Weber}},
  \bibinfo {author} {\bibfnamefont {J.~B.}\ \bibnamefont {Varley}}, \bibinfo
  {author} {\bibfnamefont {A.}~\bibnamefont {Janotti}}, \bibinfo {author}
  {\bibfnamefont {D.~D.}\ \bibnamefont {Awschalom}}, \ and\ \bibinfo {author}
  {\bibfnamefont {C.~G.}\ \bibnamefont {Van~de Walle}},\ }\href {\doibase
  10.1557/mrs.2013.206} {\bibfield  {journal} {\bibinfo  {journal} {MRS
  Bulletin}\ }\textbf {\bibinfo {volume} {38}},\ \bibinfo {pages} {802}
  (\bibinfo {year} {2013})}\BibitemShut {NoStop}%
\bibitem [{\citenamefont {Bonato}\ \emph {et~al.}(2016)\citenamefont {Bonato},
  \citenamefont {Blok}, \citenamefont {Dinani}, \citenamefont {Berry},
  \citenamefont {Markham}, \citenamefont {Twitchen},\ and\ \citenamefont
  {Hanson}}]{Bonato:2016eu}%
  \BibitemOpen
  \bibfield  {author} {\bibinfo {author} {\bibfnamefont {C.}~\bibnamefont
  {Bonato}}, \bibinfo {author} {\bibfnamefont {M.~S.}\ \bibnamefont {Blok}},
  \bibinfo {author} {\bibfnamefont {H.~T.}\ \bibnamefont {Dinani}}, \bibinfo
  {author} {\bibfnamefont {D.~W.}\ \bibnamefont {Berry}}, \bibinfo {author}
  {\bibfnamefont {M.~L.}\ \bibnamefont {Markham}}, \bibinfo {author}
  {\bibfnamefont {D.~J.}\ \bibnamefont {Twitchen}}, \ and\ \bibinfo {author}
  {\bibfnamefont {R.}~\bibnamefont {Hanson}},\ }\href {\doibase
  10.1038/nnano.2015.261} {\bibfield  {journal} {\bibinfo  {journal} {Nature
  Nanotechnology}\ }\textbf {\bibinfo {volume} {11}},\ \bibinfo {pages} {247}
  (\bibinfo {year} {2016})}\BibitemShut {NoStop}%
\bibitem [{\citenamefont {Wolfowicz}, \citenamefont {Whiteley},\ and\
  \citenamefont {Awschalom}(2018)}]{Wolfowicz:2018ts}%
  \BibitemOpen
  \bibfield  {author} {\bibinfo {author} {\bibfnamefont {G.}~\bibnamefont
  {Wolfowicz}}, \bibinfo {author} {\bibfnamefont {S.~J.}\ \bibnamefont
  {Whiteley}}, \ and\ \bibinfo {author} {\bibfnamefont {D.~D.}\ \bibnamefont
  {Awschalom}},\ }\href {https://arxiv.org/abs/1803.05956} {\  (\bibinfo {year}
  {2018})},\ \Eprint {http://arxiv.org/abs/1803.05956} {1803.05956}
  \BibitemShut {NoStop}%
\bibitem [{\citenamefont {Tarasenko}\ \emph {et~al.}(2018)\citenamefont
  {Tarasenko}, \citenamefont {Poshakinskiy}, \citenamefont {Simin},
  \citenamefont {Soltamov}, \citenamefont {Mokhov}, \citenamefont {Baranov},
  \citenamefont {Dyakonov},\ and\ \citenamefont {Astakhov}}]{Tarasenko:2018ky}%
  \BibitemOpen
  \bibfield  {author} {\bibinfo {author} {\bibfnamefont {S.~A.}\ \bibnamefont
  {Tarasenko}}, \bibinfo {author} {\bibfnamefont {A.~V.}\ \bibnamefont
  {Poshakinskiy}}, \bibinfo {author} {\bibfnamefont {D.}~\bibnamefont {Simin}},
  \bibinfo {author} {\bibfnamefont {V.~A.}\ \bibnamefont {Soltamov}}, \bibinfo
  {author} {\bibfnamefont {E.~N.}\ \bibnamefont {Mokhov}}, \bibinfo {author}
  {\bibfnamefont {P.~G.}\ \bibnamefont {Baranov}}, \bibinfo {author}
  {\bibfnamefont {V.}~\bibnamefont {Dyakonov}}, \ and\ \bibinfo {author}
  {\bibfnamefont {G.~V.}\ \bibnamefont {Astakhov}},\ }\href {\doibase
  10.1002/pssb.201700258} {\bibfield  {journal} {\bibinfo  {journal} {Physica
  Status Solidi (b)}\ }\textbf {\bibinfo {volume} {255}},\ \bibinfo {pages}
  {1700258} (\bibinfo {year} {2018})}\BibitemShut {NoStop}%
\bibitem [{\citenamefont {Blok}\ \emph {et~al.}(2015)\citenamefont {Blok},
  \citenamefont {Kalb}, \citenamefont {Reiserer}, \citenamefont {Taminiau},\
  and\ \citenamefont {Hanson}}]{Blok:2015di}%
  \BibitemOpen
  \bibfield  {author} {\bibinfo {author} {\bibfnamefont {M.~S.}\ \bibnamefont
  {Blok}}, \bibinfo {author} {\bibfnamefont {N.}~\bibnamefont {Kalb}}, \bibinfo
  {author} {\bibfnamefont {A.}~\bibnamefont {Reiserer}}, \bibinfo {author}
  {\bibfnamefont {T.~H.}\ \bibnamefont {Taminiau}}, \ and\ \bibinfo {author}
  {\bibfnamefont {R.}~\bibnamefont {Hanson}},\ }\href {\doibase
  10.1039/C5FD00113G} {\bibfield  {journal} {\bibinfo  {journal} {Faraday
  Discuss.}\ } (\bibinfo {year} {2015}),\ 10.1039/C5FD00113G}\BibitemShut
  {NoStop}%
\bibitem [{\citenamefont {Johnson}, \citenamefont {Dolan},\ and\ \citenamefont
  {Smith}(2017)}]{Johnson:2017jl}%
  \BibitemOpen
  \bibfield  {author} {\bibinfo {author} {\bibfnamefont {S.}~\bibnamefont
  {Johnson}}, \bibinfo {author} {\bibfnamefont {P.~R.}\ \bibnamefont {Dolan}},
  \ and\ \bibinfo {author} {\bibfnamefont {J.~M.}\ \bibnamefont {Smith}},\
  }\href {\doibase 10.1016/j.pquantelec.2017.05.003} {\bibfield  {journal}
  {\bibinfo  {journal} {Progress in Quantum Electronics}\ }\textbf {\bibinfo
  {volume} {55}},\ \bibinfo {pages} {129} (\bibinfo {year} {2017})}\BibitemShut
  {NoStop}%
\bibitem [{\citenamefont {Bernien}\ \emph {et~al.}(2013)\citenamefont
  {Bernien}, \citenamefont {Hensen}, \citenamefont {Pfaff}, \citenamefont
  {Koolstra}, \citenamefont {Blok}, \citenamefont {Robledo}, \citenamefont
  {Taminiau}, \citenamefont {Markham}, \citenamefont {Twitchen}, \citenamefont
  {Childress},\ and\ \citenamefont {Hanson}}]{Bernien:2013kj}%
  \BibitemOpen
  \bibfield  {author} {\bibinfo {author} {\bibfnamefont {H.}~\bibnamefont
  {Bernien}}, \bibinfo {author} {\bibfnamefont {B.}~\bibnamefont {Hensen}},
  \bibinfo {author} {\bibfnamefont {W.}~\bibnamefont {Pfaff}}, \bibinfo
  {author} {\bibfnamefont {G.}~\bibnamefont {Koolstra}}, \bibinfo {author}
  {\bibfnamefont {M.~S.}\ \bibnamefont {Blok}}, \bibinfo {author}
  {\bibfnamefont {L.}~\bibnamefont {Robledo}}, \bibinfo {author} {\bibfnamefont
  {T.~H.}\ \bibnamefont {Taminiau}}, \bibinfo {author} {\bibfnamefont
  {M.}~\bibnamefont {Markham}}, \bibinfo {author} {\bibfnamefont {D.~J.}\
  \bibnamefont {Twitchen}}, \bibinfo {author} {\bibfnamefont {L.}~\bibnamefont
  {Childress}}, \ and\ \bibinfo {author} {\bibfnamefont {R.}~\bibnamefont
  {Hanson}},\ }\href {\doibase 10.1038/nature12016} {\bibfield  {journal}
  {\bibinfo  {journal} {Nature}\ }\textbf {\bibinfo {volume} {497}},\ \bibinfo
  {pages} {86} (\bibinfo {year} {2013})}\BibitemShut {NoStop}%
\bibitem [{\citenamefont {Pfaff}\ \emph {et~al.}(2014)\citenamefont {Pfaff},
  \citenamefont {Hensen}, \citenamefont {Bernien}, \citenamefont {van Dam},
  \citenamefont {Blok}, \citenamefont {Taminiau}, \citenamefont {Tiggelman},
  \citenamefont {Schouten}, \citenamefont {Markham}, \citenamefont {Twitchen},\
  and\ \citenamefont {Hanson}}]{Pfaff:2014hy}%
  \BibitemOpen
  \bibfield  {author} {\bibinfo {author} {\bibfnamefont {W.}~\bibnamefont
  {Pfaff}}, \bibinfo {author} {\bibfnamefont {B.~J.}\ \bibnamefont {Hensen}},
  \bibinfo {author} {\bibfnamefont {H.}~\bibnamefont {Bernien}}, \bibinfo
  {author} {\bibfnamefont {S.~B.}\ \bibnamefont {van Dam}}, \bibinfo {author}
  {\bibfnamefont {M.~S.}\ \bibnamefont {Blok}}, \bibinfo {author}
  {\bibfnamefont {T.~H.}\ \bibnamefont {Taminiau}}, \bibinfo {author}
  {\bibfnamefont {M.~J.}\ \bibnamefont {Tiggelman}}, \bibinfo {author}
  {\bibfnamefont {R.~N.}\ \bibnamefont {Schouten}}, \bibinfo {author}
  {\bibfnamefont {M.}~\bibnamefont {Markham}}, \bibinfo {author} {\bibfnamefont
  {D.~J.}\ \bibnamefont {Twitchen}}, \ and\ \bibinfo {author} {\bibfnamefont
  {R.}~\bibnamefont {Hanson}},\ }\href {\doibase 10.1126/science.1253512}
  {\bibfield  {journal} {\bibinfo  {journal} {Science}\ }\textbf {\bibinfo
  {volume} {345}},\ \bibinfo {pages} {532} (\bibinfo {year}
  {2014})}\BibitemShut {NoStop}%
\bibitem [{\citenamefont {Atat{\"u}re}\ \emph {et~al.}(2018)\citenamefont
  {Atat{\"u}re}, \citenamefont {Englund}, \citenamefont {Vamivakas},
  \citenamefont {Lee},\ and\ \citenamefont {Wrachtrup}}]{Atature:2018hh}%
  \BibitemOpen
  \bibfield  {author} {\bibinfo {author} {\bibfnamefont {M.}~\bibnamefont
  {Atat{\"u}re}}, \bibinfo {author} {\bibfnamefont {D.}~\bibnamefont
  {Englund}}, \bibinfo {author} {\bibfnamefont {N.}~\bibnamefont {Vamivakas}},
  \bibinfo {author} {\bibfnamefont {S.-Y.}\ \bibnamefont {Lee}}, \ and\
  \bibinfo {author} {\bibfnamefont {J.}~\bibnamefont {Wrachtrup}},\ }\href
  {\doibase 10.1038/s41578-018-0008-9} {\bibfield  {journal} {\bibinfo
  {journal} {Nature Reviews Materials}\ }\textbf {\bibinfo {volume} {491}},\
  \bibinfo {pages} {421} (\bibinfo {year} {2018})}\BibitemShut {NoStop}%
\bibitem [{\citenamefont {Christle}\ \emph {et~al.}(2017)\citenamefont
  {Christle}, \citenamefont {Klimov}, \citenamefont {de~las Casas},
  \citenamefont {Sz{\'a}sz}, \citenamefont {Iv{\'a}dy}, \citenamefont
  {Jokubavicius}, \citenamefont {Hassan}, \citenamefont {Syv{\"a}j{\"a}rvi},
  \citenamefont {Koehl}, \citenamefont {Ohshima}, \citenamefont {Son},
  \citenamefont {Janz{\'e}n}, \citenamefont {Gali},\ and\ \citenamefont
  {Awschalom}}]{Christle:2017cy}%
  \BibitemOpen
  \bibfield  {author} {\bibinfo {author} {\bibfnamefont {D.~J.}\ \bibnamefont
  {Christle}}, \bibinfo {author} {\bibfnamefont {P.~V.}\ \bibnamefont
  {Klimov}}, \bibinfo {author} {\bibfnamefont {C.~F.}\ \bibnamefont {de~las
  Casas}}, \bibinfo {author} {\bibfnamefont {K.}~\bibnamefont {Sz{\'a}sz}},
  \bibinfo {author} {\bibfnamefont {V.}~\bibnamefont {Iv{\'a}dy}}, \bibinfo
  {author} {\bibfnamefont {V.}~\bibnamefont {Jokubavicius}}, \bibinfo {author}
  {\bibfnamefont {J.~U.}\ \bibnamefont {Hassan}}, \bibinfo {author}
  {\bibfnamefont {M.}~\bibnamefont {Syv{\"a}j{\"a}rvi}}, \bibinfo {author}
  {\bibfnamefont {W.~F.}\ \bibnamefont {Koehl}}, \bibinfo {author}
  {\bibfnamefont {T.}~\bibnamefont {Ohshima}}, \bibinfo {author} {\bibfnamefont
  {N.~T.}\ \bibnamefont {Son}}, \bibinfo {author} {\bibfnamefont
  {E.}~\bibnamefont {Janz{\'e}n}}, \bibinfo {author} {\bibfnamefont
  {{\'A}.}~\bibnamefont {Gali}}, \ and\ \bibinfo {author} {\bibfnamefont
  {D.~D.}\ \bibnamefont {Awschalom}},\ }\href {\doibase
  10.1103/PhysRevX.7.021046} {\bibfield  {journal} {\bibinfo  {journal}
  {Physical Review X}\ }\textbf {\bibinfo {volume} {7}},\ \bibinfo {pages}
  {021046} (\bibinfo {year} {2017})}\BibitemShut {NoStop}%
\bibitem [{\citenamefont {Zhong}\ \emph {et~al.}(2015)\citenamefont {Zhong},
  \citenamefont {Kindem}, \citenamefont {Miyazono},\ and\ \citenamefont
  {Faraon}}]{Zhong:2015iy}%
  \BibitemOpen
  \bibfield  {author} {\bibinfo {author} {\bibfnamefont {T.}~\bibnamefont
  {Zhong}}, \bibinfo {author} {\bibfnamefont {J.~M.}\ \bibnamefont {Kindem}},
  \bibinfo {author} {\bibfnamefont {E.}~\bibnamefont {Miyazono}}, \ and\
  \bibinfo {author} {\bibfnamefont {A.}~\bibnamefont {Faraon}},\ }\href
  {\doibase 10.1038/ncomms9206} {\bibfield  {journal} {\bibinfo  {journal}
  {Nature Communications}\ }\textbf {\bibinfo {volume} {6}},\ \bibinfo {pages}
  {8206} (\bibinfo {year} {2015})}\BibitemShut {NoStop}%
\bibitem [{\citenamefont {Steger}\ \emph {et~al.}(2012)\citenamefont {Steger},
  \citenamefont {Saeedi}, \citenamefont {Thewalt}, \citenamefont {Simmons},
  \citenamefont {Riemann}, \citenamefont {Abrosimov}, \citenamefont {Becker},\
  and\ \citenamefont {Pohl}}]{Steger:2012ev}%
  \BibitemOpen
  \bibfield  {author} {\bibinfo {author} {\bibfnamefont {M.}~\bibnamefont
  {Steger}}, \bibinfo {author} {\bibfnamefont {K.}~\bibnamefont {Saeedi}},
  \bibinfo {author} {\bibfnamefont {M.~L.~W.}\ \bibnamefont {Thewalt}},
  \bibinfo {author} {\bibfnamefont {S.}~\bibnamefont {Simmons}}, \bibinfo
  {author} {\bibfnamefont {H.}~\bibnamefont {Riemann}}, \bibinfo {author}
  {\bibfnamefont {N.~V.}\ \bibnamefont {Abrosimov}}, \bibinfo {author}
  {\bibfnamefont {P.}~\bibnamefont {Becker}}, \ and\ \bibinfo {author}
  {\bibfnamefont {H.~J.}\ \bibnamefont {Pohl}},\ }\href {\doibase
  10.1126/science.1217635} {\bibfield  {journal} {\bibinfo  {journal}
  {Science}\ }\textbf {\bibinfo {volume} {336}},\ \bibinfo {pages} {1280}
  (\bibinfo {year} {2012})}\BibitemShut {NoStop}%
\bibitem [{\citenamefont {Yang}\ \emph {et~al.}(2006)\citenamefont {Yang},
  \citenamefont {Steger}, \citenamefont {Karaiskaj}, \citenamefont {Thewalt},
  \citenamefont {Cardona}, \citenamefont {Itoh}, \citenamefont {Riemann},
  \citenamefont {Abrosimov}, \citenamefont {Churbanov}, \citenamefont {Gusev},
  \citenamefont {Bulanov}, \citenamefont {Kaliteevskii}, \citenamefont
  {Godisov}, \citenamefont {Becker}, \citenamefont {Pohl}, \citenamefont
  {Ager},\ and\ \citenamefont {Haller}}]{Yang:2006gz}%
  \BibitemOpen
  \bibfield  {author} {\bibinfo {author} {\bibfnamefont {A.}~\bibnamefont
  {Yang}}, \bibinfo {author} {\bibfnamefont {M.}~\bibnamefont {Steger}},
  \bibinfo {author} {\bibfnamefont {D.}~\bibnamefont {Karaiskaj}}, \bibinfo
  {author} {\bibfnamefont {M.~L.~W.}\ \bibnamefont {Thewalt}}, \bibinfo
  {author} {\bibfnamefont {M.}~\bibnamefont {Cardona}}, \bibinfo {author}
  {\bibfnamefont {K.~M.}\ \bibnamefont {Itoh}}, \bibinfo {author}
  {\bibfnamefont {H.}~\bibnamefont {Riemann}}, \bibinfo {author} {\bibfnamefont
  {N.~V.}\ \bibnamefont {Abrosimov}}, \bibinfo {author} {\bibfnamefont {M.~F.}\
  \bibnamefont {Churbanov}}, \bibinfo {author} {\bibfnamefont {A.~V.}\
  \bibnamefont {Gusev}}, \bibinfo {author} {\bibfnamefont {A.~D.}\ \bibnamefont
  {Bulanov}}, \bibinfo {author} {\bibfnamefont {A.~K.}\ \bibnamefont
  {Kaliteevskii}}, \bibinfo {author} {\bibfnamefont {O.~N.}\ \bibnamefont
  {Godisov}}, \bibinfo {author} {\bibfnamefont {P.}~\bibnamefont {Becker}},
  \bibinfo {author} {\bibfnamefont {H.~J.}\ \bibnamefont {Pohl}}, \bibinfo
  {author} {\bibfnamefont {J.~W.}\ \bibnamefont {Ager}}, \ and\ \bibinfo
  {author} {\bibfnamefont {E.~E.}\ \bibnamefont {Haller}},\ }\href {\doibase
  10.1103/PhysRevLett.97.227401} {\bibfield  {journal} {\bibinfo  {journal}
  {Physical Review Letters}\ }\textbf {\bibinfo {volume} {97}},\ \bibinfo
  {pages} {227401} (\bibinfo {year} {2006})}\BibitemShut {NoStop}%
\bibitem [{\citenamefont {Thewalt}\ \emph {et~al.}(2007)\citenamefont
  {Thewalt}, \citenamefont {Yang}, \citenamefont {Steger}, \citenamefont
  {Karaiskaj}, \citenamefont {Cardona}, \citenamefont {Riemann}, \citenamefont
  {Abrosimov}, \citenamefont {Gusev}, \citenamefont {Bulanov}, \citenamefont
  {Kovalev}, \citenamefont {Kaliteevskii}, \citenamefont {Godisov},
  \citenamefont {Becker}, \citenamefont {Pohl}, \citenamefont {Haller},
  \citenamefont {Ager~Iii},\ and\ \citenamefont {Itoh}}]{Thewalt:2007bq}%
  \BibitemOpen
  \bibfield  {author} {\bibinfo {author} {\bibfnamefont {M.~L.~W.}\
  \bibnamefont {Thewalt}}, \bibinfo {author} {\bibfnamefont {A.}~\bibnamefont
  {Yang}}, \bibinfo {author} {\bibfnamefont {M.}~\bibnamefont {Steger}},
  \bibinfo {author} {\bibfnamefont {D.}~\bibnamefont {Karaiskaj}}, \bibinfo
  {author} {\bibfnamefont {M.}~\bibnamefont {Cardona}}, \bibinfo {author}
  {\bibfnamefont {H.}~\bibnamefont {Riemann}}, \bibinfo {author} {\bibfnamefont
  {N.~V.}\ \bibnamefont {Abrosimov}}, \bibinfo {author} {\bibfnamefont {A.~V.}\
  \bibnamefont {Gusev}}, \bibinfo {author} {\bibfnamefont {A.~D.}\ \bibnamefont
  {Bulanov}}, \bibinfo {author} {\bibfnamefont {I.~D.}\ \bibnamefont
  {Kovalev}}, \bibinfo {author} {\bibfnamefont {A.~K.}\ \bibnamefont
  {Kaliteevskii}}, \bibinfo {author} {\bibfnamefont {O.~N.}\ \bibnamefont
  {Godisov}}, \bibinfo {author} {\bibfnamefont {P.}~\bibnamefont {Becker}},
  \bibinfo {author} {\bibfnamefont {H.~J.}\ \bibnamefont {Pohl}}, \bibinfo
  {author} {\bibfnamefont {E.~E.}\ \bibnamefont {Haller}}, \bibinfo {author}
  {\bibfnamefont {J.~W.}\ \bibnamefont {Ager~Iii}}, \ and\ \bibinfo {author}
  {\bibfnamefont {K.~M.}\ \bibnamefont {Itoh}},\ }\href {\doibase
  10.1063/1.2723181} {\bibfield  {journal} {\bibinfo  {journal} {Journal of
  Applied Physics}\ }\textbf {\bibinfo {volume} {101}},\ \bibinfo {pages}
  {081724} (\bibinfo {year} {2007})}\BibitemShut {NoStop}%
\bibitem [{\citenamefont {Steger}\ \emph {et~al.}(2011)\citenamefont {Steger},
  \citenamefont {Sekiguchi}, \citenamefont {Yang}, \citenamefont {Saeedi},
  \citenamefont {Hayden}, \citenamefont {Thewalt}, \citenamefont {Itoh},
  \citenamefont {Riemann}, \citenamefont {Abrosimov}, \citenamefont {Becker},\
  and\ \citenamefont {Pohl}}]{Steger:2011fh}%
  \BibitemOpen
  \bibfield  {author} {\bibinfo {author} {\bibfnamefont {M.}~\bibnamefont
  {Steger}}, \bibinfo {author} {\bibfnamefont {T.}~\bibnamefont {Sekiguchi}},
  \bibinfo {author} {\bibfnamefont {A.}~\bibnamefont {Yang}}, \bibinfo {author}
  {\bibfnamefont {K.}~\bibnamefont {Saeedi}}, \bibinfo {author} {\bibfnamefont
  {M.~E.}\ \bibnamefont {Hayden}}, \bibinfo {author} {\bibfnamefont {M.~L.~W.}\
  \bibnamefont {Thewalt}}, \bibinfo {author} {\bibfnamefont {K.~M.}\
  \bibnamefont {Itoh}}, \bibinfo {author} {\bibfnamefont {H.}~\bibnamefont
  {Riemann}}, \bibinfo {author} {\bibfnamefont {N.~V.}\ \bibnamefont
  {Abrosimov}}, \bibinfo {author} {\bibfnamefont {P.}~\bibnamefont {Becker}}, \
  and\ \bibinfo {author} {\bibfnamefont {H.~J.}\ \bibnamefont {Pohl}},\ }\href
  {\doibase 10.1063/1.3577614} {\bibfield  {journal} {\bibinfo  {journal}
  {Journal of Applied Physics}\ }\textbf {\bibinfo {volume} {109}},\ \bibinfo
  {pages} {102411} (\bibinfo {year} {2011})}\BibitemShut {NoStop}%
\bibitem [{\citenamefont {Zwier}\ \emph {et~al.}(2015)\citenamefont {Zwier},
  \citenamefont {O{\textquoteright}Shea}, \citenamefont {Onur},\ and\
  \citenamefont {van~der Wal}}]{Zwier:2015go}%
  \BibitemOpen
  \bibfield  {author} {\bibinfo {author} {\bibfnamefont {O.~V.}\ \bibnamefont
  {Zwier}}, \bibinfo {author} {\bibfnamefont {D.}~\bibnamefont
  {O{\textquoteright}Shea}}, \bibinfo {author} {\bibfnamefont {A.~R.}\
  \bibnamefont {Onur}}, \ and\ \bibinfo {author} {\bibfnamefont {C.~H.}\
  \bibnamefont {van~der Wal}},\ }\href {\doibase 10.1038/srep10931} {\bibfield
  {journal} {\bibinfo  {journal} {Scientific Reports}\ }\textbf {\bibinfo
  {volume} {5}},\ \bibinfo {pages} {10931} (\bibinfo {year}
  {2015})}\BibitemShut {NoStop}%
\bibitem [{\citenamefont {Schmid}(1977)}]{Schmid:1977bi}%
  \BibitemOpen
  \bibfield  {author} {\bibinfo {author} {\bibfnamefont {W.}~\bibnamefont
  {Schmid}},\ }\href {\doibase 10.1002/pssb.2220840216} {\bibfield  {journal}
  {\bibinfo  {journal} {Physica Status Solidi (b)}\ }\textbf {\bibinfo {volume}
  {84}},\ \bibinfo {pages} {529} (\bibinfo {year} {1977})}\BibitemShut
  {NoStop}%
\bibitem [{\citenamefont {Sumikura}\ \emph {et~al.}(2011)\citenamefont
  {Sumikura}, \citenamefont {Nishiguchi}, \citenamefont {Ono}, \citenamefont
  {Fujiwara},\ and\ \citenamefont {Notomi}}]{Sumikura:2011ch}%
  \BibitemOpen
  \bibfield  {author} {\bibinfo {author} {\bibfnamefont {H.}~\bibnamefont
  {Sumikura}}, \bibinfo {author} {\bibfnamefont {K.}~\bibnamefont
  {Nishiguchi}}, \bibinfo {author} {\bibfnamefont {Y.}~\bibnamefont {Ono}},
  \bibinfo {author} {\bibfnamefont {A.}~\bibnamefont {Fujiwara}}, \ and\
  \bibinfo {author} {\bibfnamefont {M.}~\bibnamefont {Notomi}},\ }\href
  {\doibase 10.1364/OE.19.025255} {\bibfield  {journal} {\bibinfo  {journal}
  {Optics Express}\ }\textbf {\bibinfo {volume} {19}},\ \bibinfo {pages}
  {25255} (\bibinfo {year} {2011})}\BibitemShut {NoStop}%
\bibitem [{\citenamefont {Pelton}\ \emph {et~al.}(2002)\citenamefont {Pelton},
  \citenamefont {Santori}, \citenamefont {Vuckovic}, \citenamefont {Zhang},
  \citenamefont {Solomon}, \citenamefont {Plant},\ and\ \citenamefont
  {Yamamoto}}]{Pelton:2002hv}%
  \BibitemOpen
  \bibfield  {author} {\bibinfo {author} {\bibfnamefont {M.}~\bibnamefont
  {Pelton}}, \bibinfo {author} {\bibfnamefont {C.}~\bibnamefont {Santori}},
  \bibinfo {author} {\bibfnamefont {J.}~\bibnamefont {Vuckovic}}, \bibinfo
  {author} {\bibfnamefont {B.}~\bibnamefont {Zhang}}, \bibinfo {author}
  {\bibfnamefont {G.~S.}\ \bibnamefont {Solomon}}, \bibinfo {author}
  {\bibfnamefont {J.}~\bibnamefont {Plant}}, \ and\ \bibinfo {author}
  {\bibfnamefont {Y.}~\bibnamefont {Yamamoto}},\ }\href {\doibase
  10.1103/PhysRevLett.89.233602} {\bibfield  {journal} {\bibinfo  {journal}
  {Physical Review Letters}\ }\textbf {\bibinfo {volume} {89}},\ \bibinfo
  {pages} {233602} (\bibinfo {year} {2002})}\BibitemShut {NoStop}%
\bibitem [{\citenamefont {Davan{\c c}o}\ \emph {et~al.}(2011)\citenamefont
  {Davan{\c c}o}, \citenamefont {Rakher}, \citenamefont {Schuh}, \citenamefont
  {Badolato},\ and\ \citenamefont {Srinivasan}}]{Davanco:2011iv}%
  \BibitemOpen
  \bibfield  {author} {\bibinfo {author} {\bibfnamefont {M.}~\bibnamefont
  {Davan{\c c}o}}, \bibinfo {author} {\bibfnamefont {M.~T.}\ \bibnamefont
  {Rakher}}, \bibinfo {author} {\bibfnamefont {D.}~\bibnamefont {Schuh}},
  \bibinfo {author} {\bibfnamefont {A.}~\bibnamefont {Badolato}}, \ and\
  \bibinfo {author} {\bibfnamefont {K.}~\bibnamefont {Srinivasan}},\ }\href
  {\doibase 10.1063/1.3615051} {\bibfield  {journal} {\bibinfo  {journal}
  {Applied Physics Letters}\ }\textbf {\bibinfo {volume} {99}},\ \bibinfo
  {pages} {041102} (\bibinfo {year} {2011})}\BibitemShut {NoStop}%
\bibitem [{\citenamefont {Joannopoulos}, \citenamefont {Villeneuve},\ and\
  \citenamefont {Fan}(1997)}]{Joannopoulos:1997ez}%
  \BibitemOpen
  \bibfield  {author} {\bibinfo {author} {\bibfnamefont {J.~D.}\ \bibnamefont
  {Joannopoulos}}, \bibinfo {author} {\bibfnamefont {P.~R.}\ \bibnamefont
  {Villeneuve}}, \ and\ \bibinfo {author} {\bibfnamefont {S.}~\bibnamefont
  {Fan}},\ }\href {\doibase 10.1038/386143a0} {\bibfield  {journal} {\bibinfo
  {journal} {Nature}\ }\textbf {\bibinfo {volume} {386}},\ \bibinfo {pages}
  {143} (\bibinfo {year} {1997})}\BibitemShut {NoStop}%
\bibitem [{\citenamefont {Faraon}\ \emph {et~al.}(2012)\citenamefont {Faraon},
  \citenamefont {Santori}, \citenamefont {Huang}, \citenamefont {Acosta},\ and\
  \citenamefont {Beausoleil}}]{Faraon:2012ky}%
  \BibitemOpen
  \bibfield  {author} {\bibinfo {author} {\bibfnamefont {A.}~\bibnamefont
  {Faraon}}, \bibinfo {author} {\bibfnamefont {C.}~\bibnamefont {Santori}},
  \bibinfo {author} {\bibfnamefont {Z.}~\bibnamefont {Huang}}, \bibinfo
  {author} {\bibfnamefont {V.~M.}\ \bibnamefont {Acosta}}, \ and\ \bibinfo
  {author} {\bibfnamefont {R.~G.}\ \bibnamefont {Beausoleil}},\ }\href
  {\doibase 10.1103/PhysRevLett.109.033604} {\bibfield  {journal} {\bibinfo
  {journal} {Physical Review Letters}\ }\textbf {\bibinfo {volume} {109}},\
  \bibinfo {pages} {033604} (\bibinfo {year} {2012})}\BibitemShut {NoStop}%
\bibitem [{\citenamefont {Reinhard}\ \emph {et~al.}(2011)\citenamefont
  {Reinhard}, \citenamefont {Volz}, \citenamefont {Winger}, \citenamefont
  {Badolato}, \citenamefont {Hennessy}, \citenamefont {Hu},\ and\ \citenamefont
  {Imamoglu}}]{Reinhard:2011er}%
  \BibitemOpen
  \bibfield  {author} {\bibinfo {author} {\bibfnamefont {A.}~\bibnamefont
  {Reinhard}}, \bibinfo {author} {\bibfnamefont {T.}~\bibnamefont {Volz}},
  \bibinfo {author} {\bibfnamefont {M.}~\bibnamefont {Winger}}, \bibinfo
  {author} {\bibfnamefont {A.}~\bibnamefont {Badolato}}, \bibinfo {author}
  {\bibfnamefont {K.~J.}\ \bibnamefont {Hennessy}}, \bibinfo {author}
  {\bibfnamefont {E.~L.}\ \bibnamefont {Hu}}, \ and\ \bibinfo {author}
  {\bibfnamefont {A.}~\bibnamefont {Imamoglu}},\ }\href {\doibase
  10.1038/nphoton.2011.321} {\bibfield  {journal} {\bibinfo  {journal} {Nature
  Photonics}\ }\textbf {\bibinfo {volume} {6}},\ \bibinfo {pages} {93}
  (\bibinfo {year} {2011})}\BibitemShut {NoStop}%
\bibitem [{\citenamefont {Lee}\ \emph {et~al.}(2015)\citenamefont {Lee},
  \citenamefont {Lim}, \citenamefont {Schneider}, \citenamefont {Maier},
  \citenamefont {H{\"o}fling}, \citenamefont {Kamp},\ and\ \citenamefont
  {Lee}}]{Lee:2015ig}%
  \BibitemOpen
  \bibfield  {author} {\bibinfo {author} {\bibfnamefont {C.-M.}\ \bibnamefont
  {Lee}}, \bibinfo {author} {\bibfnamefont {H.-J.}\ \bibnamefont {Lim}},
  \bibinfo {author} {\bibfnamefont {C.}~\bibnamefont {Schneider}}, \bibinfo
  {author} {\bibfnamefont {S.}~\bibnamefont {Maier}}, \bibinfo {author}
  {\bibfnamefont {S.}~\bibnamefont {H{\"o}fling}}, \bibinfo {author}
  {\bibfnamefont {M.}~\bibnamefont {Kamp}}, \ and\ \bibinfo {author}
  {\bibfnamefont {Y.-H.}\ \bibnamefont {Lee}},\ }\href {\doibase
  10.1038/srep14309} {\bibfield  {journal} {\bibinfo  {journal} {Sci. Rep.}\
  }\textbf {\bibinfo {volume} {5}},\ \bibinfo {pages} {14309} (\bibinfo {year}
  {2015})}\BibitemShut {NoStop}%
\bibitem [{\citenamefont {Fu}\ \emph {et~al.}(2004)\citenamefont {Fu},
  \citenamefont {Ladd}, \citenamefont {Santori},\ and\ \citenamefont
  {Yamamoto}}]{Fu:2004kp}%
  \BibitemOpen
  \bibfield  {author} {\bibinfo {author} {\bibfnamefont {K.-M.~C.}\
  \bibnamefont {Fu}}, \bibinfo {author} {\bibfnamefont {T.~D.}\ \bibnamefont
  {Ladd}}, \bibinfo {author} {\bibfnamefont {C.}~\bibnamefont {Santori}}, \
  and\ \bibinfo {author} {\bibfnamefont {Y.}~\bibnamefont {Yamamoto}},\ }\href
  {\doibase 10.1103/PhysRevB.69.125306} {\bibfield  {journal} {\bibinfo
  {journal} {Physical Review B}\ }\textbf {\bibinfo {volume} {69}},\ \bibinfo
  {pages} {125306} (\bibinfo {year} {2004})}\BibitemShut {NoStop}%
\bibitem [{\citenamefont {Calusine}, \citenamefont {Politi},\ and\
  \citenamefont {Awschalom}(2014)}]{Calusine:2014gv}%
  \BibitemOpen
  \bibfield  {author} {\bibinfo {author} {\bibfnamefont {G.}~\bibnamefont
  {Calusine}}, \bibinfo {author} {\bibfnamefont {A.}~\bibnamefont {Politi}}, \
  and\ \bibinfo {author} {\bibfnamefont {D.~D.}\ \bibnamefont {Awschalom}},\
  }\href {\doibase 10.1063/1.4890083} {\bibfield  {journal} {\bibinfo
  {journal} {Applied Physics Letters}\ }\textbf {\bibinfo {volume} {105}},\
  \bibinfo {pages} {011123} (\bibinfo {year} {2014})}\BibitemShut {NoStop}%
\bibitem [{\citenamefont {Calusine}, \citenamefont {Politi},\ and\
  \citenamefont {Awschalom}(2016)}]{Calusine:2016hr}%
  \BibitemOpen
  \bibfield  {author} {\bibinfo {author} {\bibfnamefont {G.}~\bibnamefont
  {Calusine}}, \bibinfo {author} {\bibfnamefont {A.}~\bibnamefont {Politi}}, \
  and\ \bibinfo {author} {\bibfnamefont {D.~D.}\ \bibnamefont {Awschalom}},\
  }\href {\doibase 10.1103/PhysRevApplied.6.014019} {\bibfield  {journal}
  {\bibinfo  {journal} {Physical Review Applied}\ }\textbf {\bibinfo {volume}
  {6}},\ \bibinfo {pages} {014019} (\bibinfo {year} {2016})}\BibitemShut
  {NoStop}%
\bibitem [{\citenamefont {Dean}, \citenamefont {Haynes},\ and\ \citenamefont
  {Flood}(1967)}]{Dean:1967dl}%
  \BibitemOpen
  \bibfield  {author} {\bibinfo {author} {\bibfnamefont {P.~J.}\ \bibnamefont
  {Dean}}, \bibinfo {author} {\bibfnamefont {J.~R.}\ \bibnamefont {Haynes}}, \
  and\ \bibinfo {author} {\bibfnamefont {W.~F.}\ \bibnamefont {Flood}},\ }\href
  {\doibase 10.1103/PhysRev.161.711} {\bibfield  {journal} {\bibinfo  {journal}
  {Physical Review}\ }\textbf {\bibinfo {volume} {161}},\ \bibinfo {pages}
  {711} (\bibinfo {year} {1967})}\BibitemShut {NoStop}%
\bibitem [{\citenamefont {Lohrmann}\ \emph {et~al.}(2017)\citenamefont
  {Lohrmann}, \citenamefont {Johnson}, \citenamefont {McCallum},\ and\
  \citenamefont {Castelletto}}]{Lohrmann:2017kz}%
  \BibitemOpen
  \bibfield  {author} {\bibinfo {author} {\bibfnamefont {A.}~\bibnamefont
  {Lohrmann}}, \bibinfo {author} {\bibfnamefont {B.~C.}\ \bibnamefont
  {Johnson}}, \bibinfo {author} {\bibfnamefont {J.~C.}\ \bibnamefont
  {McCallum}}, \ and\ \bibinfo {author} {\bibfnamefont {S.}~\bibnamefont
  {Castelletto}},\ }\href {\doibase 10.1088/1361-6633/aa5171} {\bibfield
  {journal} {\bibinfo  {journal} {Reports on Progress in Physics}\ }\textbf
  {\bibinfo {volume} {80}},\ \bibinfo {pages} {034502} (\bibinfo {year}
  {2017})}\BibitemShut {NoStop}%
\bibitem [{\citenamefont {Baur}, \citenamefont {Kunzer},\ and\ \citenamefont
  {Schneider}(1997)}]{Baur:1997cx}%
  \BibitemOpen
  \bibfield  {author} {\bibinfo {author} {\bibfnamefont {J.}~\bibnamefont
  {Baur}}, \bibinfo {author} {\bibfnamefont {M.}~\bibnamefont {Kunzer}}, \ and\
  \bibinfo {author} {\bibfnamefont {J.}~\bibnamefont {Schneider}},\ }\href
  {\doibase 10.1002/1521-396X(199707)162:1<153::AID-PSSA153>3.0.CO;2-3}
  {\bibfield  {journal} {\bibinfo  {journal} {Physica Status Solidi (a)}\
  }\textbf {\bibinfo {volume} {162}},\ \bibinfo {pages} {153} (\bibinfo {year}
  {1997})}\BibitemShut {NoStop}%
\bibitem [{\citenamefont {Son}\ \emph {et~al.}(1999)\citenamefont {Son},
  \citenamefont {Ellison}, \citenamefont {Magnusson}, \citenamefont
  {MacMillan}, \citenamefont {Chen}, \citenamefont {Monemar},\ and\
  \citenamefont {Janz{\'e}n}}]{Son:1999en}%
  \BibitemOpen
  \bibfield  {author} {\bibinfo {author} {\bibfnamefont {N.~T.}\ \bibnamefont
  {Son}}, \bibinfo {author} {\bibfnamefont {A.}~\bibnamefont {Ellison}},
  \bibinfo {author} {\bibfnamefont {B.}~\bibnamefont {Magnusson}}, \bibinfo
  {author} {\bibfnamefont {M.~F.}\ \bibnamefont {MacMillan}}, \bibinfo {author}
  {\bibfnamefont {W.~M.}\ \bibnamefont {Chen}}, \bibinfo {author}
  {\bibfnamefont {B.}~\bibnamefont {Monemar}}, \ and\ \bibinfo {author}
  {\bibfnamefont {E.}~\bibnamefont {Janz{\'e}n}},\ }\href {\doibase
  10.1063/1.371368} {\bibfield  {journal} {\bibinfo  {journal} {Journal of
  Applied Physics}\ }\textbf {\bibinfo {volume} {86}},\ \bibinfo {pages} {4348}
  (\bibinfo {year} {1999})}\BibitemShut {NoStop}%
\bibitem [{\citenamefont {Son}\ \emph {et~al.}(1996)\citenamefont {Son},
  \citenamefont {S{\"o}rman}, \citenamefont {Chen}, \citenamefont {Singh},
  \citenamefont {Hallin}, \citenamefont {Kordina}, \citenamefont {Monemar},
  \citenamefont {Janz{\'e}n},\ and\ \citenamefont
  {Lindstr{\"o}m}}]{Son:1996cg}%
  \BibitemOpen
  \bibfield  {author} {\bibinfo {author} {\bibfnamefont {N.~T.}\ \bibnamefont
  {Son}}, \bibinfo {author} {\bibfnamefont {E.}~\bibnamefont {S{\"o}rman}},
  \bibinfo {author} {\bibfnamefont {W.~M.}\ \bibnamefont {Chen}}, \bibinfo
  {author} {\bibfnamefont {M.}~\bibnamefont {Singh}}, \bibinfo {author}
  {\bibfnamefont {C.}~\bibnamefont {Hallin}}, \bibinfo {author} {\bibfnamefont
  {O.}~\bibnamefont {Kordina}}, \bibinfo {author} {\bibfnamefont
  {B.}~\bibnamefont {Monemar}}, \bibinfo {author} {\bibfnamefont
  {E.}~\bibnamefont {Janz{\'e}n}}, \ and\ \bibinfo {author} {\bibfnamefont
  {J.~L.}\ \bibnamefont {Lindstr{\"o}m}},\ }\href {\doibase 10.1063/1.361214}
  {\bibfield  {journal} {\bibinfo  {journal} {Journal of Applied Physics}\
  }\textbf {\bibinfo {volume} {79}},\ \bibinfo {pages} {3784} (\bibinfo {year}
  {1996})}\BibitemShut {NoStop}%
\bibitem [{\citenamefont {Sakoda}(2005)}]{Sakoda:2005fr}%
  \BibitemOpen
  \bibfield  {author} {\bibinfo {author} {\bibfnamefont {K.}~\bibnamefont
  {Sakoda}},\ }\href {\doibase 10.1007/b138376} {\emph {\bibinfo {title}
  {{Optical Properties of Photonic Crystals}}}},\ \bibinfo {series} {Springer
  Series in Optical Sciences}, Vol.~\bibinfo {volume} {80}\ (\bibinfo
  {publisher} {Springer-Verlag},\ \bibinfo {address} {Berlin/Heidelberg},\
  \bibinfo {year} {2005})\BibitemShut {NoStop}%
\bibitem [{\citenamefont {Dussault}\ and\ \citenamefont
  {Hoess}(2004)}]{Dussault:2004fu}%
  \BibitemOpen
  \bibfield  {author} {\bibinfo {author} {\bibfnamefont {D.}~\bibnamefont
  {Dussault}}\ and\ \bibinfo {author} {\bibfnamefont {P.}~\bibnamefont
  {Hoess}},\ }in\ \href {\doibase 10.1117/12.561839} {\emph {\bibinfo
  {booktitle} {Optical Science and Technology, the SPIE 49th Annual
  Meeting}}},\ \bibinfo {editor} {edited by\ \bibinfo {editor} {\bibfnamefont
  {E.~L.}\ \bibnamefont {Dereniak}}, \bibinfo {editor} {\bibfnamefont {R.~E.}\
  \bibnamefont {Sampson}}, \ and\ \bibinfo {editor} {\bibfnamefont {C.~B.}\
  \bibnamefont {Johnson}}}\ (\bibinfo  {publisher} {SPIE},\ \bibinfo {year}
  {2004})\ p.\ \bibinfo {pages} {195}\BibitemShut {NoStop}%
\bibitem [{\citenamefont {Faraon}\ \emph {et~al.}(2008)\citenamefont {Faraon},
  \citenamefont {Fushman}, \citenamefont {Englund}, \citenamefont {Stoltz},
  \citenamefont {Petroff},\ and\ \citenamefont {Vuckovic}}]{Faraon:2008da}%
  \BibitemOpen
  \bibfield  {author} {\bibinfo {author} {\bibfnamefont {A.}~\bibnamefont
  {Faraon}}, \bibinfo {author} {\bibfnamefont {I.}~\bibnamefont {Fushman}},
  \bibinfo {author} {\bibfnamefont {D.}~\bibnamefont {Englund}}, \bibinfo
  {author} {\bibfnamefont {N.}~\bibnamefont {Stoltz}}, \bibinfo {author}
  {\bibfnamefont {P.}~\bibnamefont {Petroff}}, \ and\ \bibinfo {author}
  {\bibfnamefont {J.}~\bibnamefont {Vuckovic}},\ }\href {\doibase
  10.1038/nphys1078} {\bibfield  {journal} {\bibinfo  {journal} {Nature
  Physics}\ }\textbf {\bibinfo {volume} {4}},\ \bibinfo {pages} {859} (\bibinfo
  {year} {2008})}\BibitemShut {NoStop}%
\bibitem [{\citenamefont {Galli}\ \emph {et~al.}(2009)\citenamefont {Galli},
  \citenamefont {Portalupi}, \citenamefont {Belotti}, \citenamefont {Andreani},
  \citenamefont {O'Faolain},\ and\ \citenamefont {Krauss}}]{Galli:2009ge}%
  \BibitemOpen
  \bibfield  {author} {\bibinfo {author} {\bibfnamefont {M.}~\bibnamefont
  {Galli}}, \bibinfo {author} {\bibfnamefont {S.~L.}\ \bibnamefont
  {Portalupi}}, \bibinfo {author} {\bibfnamefont {M.}~\bibnamefont {Belotti}},
  \bibinfo {author} {\bibfnamefont {L.~C.}\ \bibnamefont {Andreani}}, \bibinfo
  {author} {\bibfnamefont {L.}~\bibnamefont {O'Faolain}}, \ and\ \bibinfo
  {author} {\bibfnamefont {T.~F.}\ \bibnamefont {Krauss}},\ }\href {\doibase
  10.1063/1.3080683} {\bibfield  {journal} {\bibinfo  {journal} {Applied
  Physics Letters}\ }\textbf {\bibinfo {volume} {94}},\ \bibinfo {pages}
  {071101} (\bibinfo {year} {2009})}\BibitemShut {NoStop}%
\bibitem [{\citenamefont {Demarest}, \citenamefont {Huang},\ and\ \citenamefont
  {Plumb}(1996)}]{Demarest:1996cu}%
  \BibitemOpen
  \bibfield  {author} {\bibinfo {author} {\bibfnamefont {K.}~\bibnamefont
  {Demarest}}, \bibinfo {author} {\bibfnamefont {Z.}~\bibnamefont {Huang}}, \
  and\ \bibinfo {author} {\bibfnamefont {R.}~\bibnamefont {Plumb}},\ }\href
  {\doibase 10.1109/8.511824} {\bibfield  {journal} {\bibinfo  {journal} {IEEE
  Transactions on Antennas and Propagation}\ }\textbf {\bibinfo {volume}
  {44}},\ \bibinfo {pages} {1150} (\bibinfo {year} {1996})}\BibitemShut
  {NoStop}%
\bibitem [{\citenamefont {Taflove}\ and\ \citenamefont
  {Hagness}(2004)}]{Taflove:2004wc}%
  \BibitemOpen
  \bibfield  {author} {\bibinfo {author} {\bibfnamefont {A.}~\bibnamefont
  {Taflove}}\ and\ \bibinfo {author} {\bibfnamefont {S.~C.}\ \bibnamefont
  {Hagness}},\ }\href
  {http://www.worldcat.org/title/orthogonal-frequency-division-multiplexing-for-wireless-communications/oclc/804515173}
  {\emph {\bibinfo {title} {{Computational Electrodynamics: The
  Finite-Difference Time-Domain Method}}}},\ \bibinfo {edition} {3rd}\ ed.\
  (\bibinfo  {publisher} {Artech House},\ \bibinfo {year} {2004})\BibitemShut
  {NoStop}%
\bibitem [{\citenamefont {Chalcraft}\ \emph {et~al.}(2007)\citenamefont
  {Chalcraft}, \citenamefont {Lam}, \citenamefont {O{\textquoteright}Brien},
  \citenamefont {Krauss}, \citenamefont {Sahin}, \citenamefont {Szymanski},
  \citenamefont {Sanvitto}, \citenamefont {Oulton}, \citenamefont {Skolnick},
  \citenamefont {Fox}, \citenamefont {Whittaker}, \citenamefont {Liu},\ and\
  \citenamefont {Hopkinson}}]{Chalcraft:2007hu}%
  \BibitemOpen
  \bibfield  {author} {\bibinfo {author} {\bibfnamefont {A.~R.~A.}\
  \bibnamefont {Chalcraft}}, \bibinfo {author} {\bibfnamefont {S.}~\bibnamefont
  {Lam}}, \bibinfo {author} {\bibfnamefont {D.}~\bibnamefont
  {O{\textquoteright}Brien}}, \bibinfo {author} {\bibfnamefont {T.~F.}\
  \bibnamefont {Krauss}}, \bibinfo {author} {\bibfnamefont {M.}~\bibnamefont
  {Sahin}}, \bibinfo {author} {\bibfnamefont {D.}~\bibnamefont {Szymanski}},
  \bibinfo {author} {\bibfnamefont {D.}~\bibnamefont {Sanvitto}}, \bibinfo
  {author} {\bibfnamefont {R.}~\bibnamefont {Oulton}}, \bibinfo {author}
  {\bibfnamefont {M.~S.}\ \bibnamefont {Skolnick}}, \bibinfo {author}
  {\bibfnamefont {A.~M.}\ \bibnamefont {Fox}}, \bibinfo {author} {\bibfnamefont
  {D.~M.}\ \bibnamefont {Whittaker}}, \bibinfo {author} {\bibfnamefont {H.~Y.}\
  \bibnamefont {Liu}}, \ and\ \bibinfo {author} {\bibfnamefont
  {M.}~\bibnamefont {Hopkinson}},\ }\href {\doibase 10.1063/1.2748310}
  {\bibfield  {journal} {\bibinfo  {journal} {Applied Physics Letters}\
  }\textbf {\bibinfo {volume} {90}},\ \bibinfo {pages} {241117} (\bibinfo
  {year} {2007})}\BibitemShut {NoStop}%
\bibitem [{\citenamefont {Akahane}\ \emph {et~al.}(2005)\citenamefont
  {Akahane}, \citenamefont {Asano}, \citenamefont {Song},\ and\ \citenamefont
  {Noda}}]{Akahane:2005ge}%
  \BibitemOpen
  \bibfield  {author} {\bibinfo {author} {\bibfnamefont {Y.}~\bibnamefont
  {Akahane}}, \bibinfo {author} {\bibfnamefont {T.}~\bibnamefont {Asano}},
  \bibinfo {author} {\bibfnamefont {B.-S.}\ \bibnamefont {Song}}, \ and\
  \bibinfo {author} {\bibfnamefont {S.}~\bibnamefont {Noda}},\ }\href {\doibase
  10.1364/OPEX.13.001202} {\bibfield  {journal} {\bibinfo  {journal} {Optics
  Express}\ }\textbf {\bibinfo {volume} {13}},\ \bibinfo {pages} {1202}
  (\bibinfo {year} {2005})}\BibitemShut {NoStop}%
\bibitem [{\citenamefont {Tran}, \citenamefont {Combri{\'e}},\ and\
  \citenamefont {De~Rossi}(2009)}]{Tran:2009vd}%
  \BibitemOpen
  \bibfield  {author} {\bibinfo {author} {\bibfnamefont {N.-V.-Q.}\
  \bibnamefont {Tran}}, \bibinfo {author} {\bibfnamefont {S.}~\bibnamefont
  {Combri{\'e}}}, \ and\ \bibinfo {author} {\bibfnamefont {A.}~\bibnamefont
  {De~Rossi}},\ }\href@noop {} {\bibfield  {journal} {\bibinfo  {journal}
  {Physical Review B}\ }\textbf {\bibinfo {volume} {79}},\ \bibinfo {pages}
  {041101} (\bibinfo {year} {2009})}\BibitemShut {NoStop}%
\bibitem [{\citenamefont {Portalupi}\ \emph {et~al.}(2010)\citenamefont
  {Portalupi}, \citenamefont {Galli}, \citenamefont {Reardon},\ and\
  \citenamefont {Krauss}}]{Portalupi:2010tr}%
  \BibitemOpen
  \bibfield  {author} {\bibinfo {author} {\bibfnamefont {S.~L.}\ \bibnamefont
  {Portalupi}}, \bibinfo {author} {\bibfnamefont {M.}~\bibnamefont {Galli}},
  \bibinfo {author} {\bibfnamefont {C.}~\bibnamefont {Reardon}}, \ and\
  \bibinfo {author} {\bibfnamefont {T.}~\bibnamefont {Krauss}},\ }\href
  {\doibase 10.1364/OE.18.016064} {\bibfield  {journal} {\bibinfo  {journal}
  {Optics Express}\ }\textbf {\bibinfo {volume} {18}},\ \bibinfo {pages}
  {16064} (\bibinfo {year} {2010})}\BibitemShut {NoStop}%
\bibitem [{\citenamefont {Bailey~III}\ \emph {et~al.}(1998)\citenamefont
  {Bailey~III}, \citenamefont {van~de Sanden}, \citenamefont {Gregus},\ and\
  \citenamefont {Gottscho}}]{BaileyIII:1998da}%
  \BibitemOpen
  \bibfield  {author} {\bibinfo {author} {\bibfnamefont {A.~D.}\ \bibnamefont
  {Bailey~III}}, \bibinfo {author} {\bibfnamefont {M.~C.~M.}\ \bibnamefont
  {van~de Sanden}}, \bibinfo {author} {\bibfnamefont {J.~A.}\ \bibnamefont
  {Gregus}}, \ and\ \bibinfo {author} {\bibfnamefont {R.~A.}\ \bibnamefont
  {Gottscho}},\ }\href {\doibase 10.1116/1.587992} {\bibfield  {journal}
  {\bibinfo  {journal} {Journal of Vacuum Science {\&} Technology B:
  Microelectronics and Nanometer Structures}\ }\textbf {\bibinfo {volume}
  {13}},\ \bibinfo {pages} {92} (\bibinfo {year} {1998})}\BibitemShut {NoStop}%
\bibitem [{\citenamefont {Novotny}\ and\ \citenamefont
  {Hecht}(2006)}]{Novotny:2006ex}%
  \BibitemOpen
  \bibfield  {author} {\bibinfo {author} {\bibfnamefont {L.}~\bibnamefont
  {Novotny}}\ and\ \bibinfo {author} {\bibfnamefont {B.}~\bibnamefont
  {Hecht}},\ }\href {\doibase 10.1017/CBO9780511813535} {\emph {\bibinfo
  {title} {{Principles of Nano-Optics}}}}\ (\bibinfo  {publisher} {Cambridge
  University Press},\ \bibinfo {address} {Cambridge},\ \bibinfo {year}
  {2006})\BibitemShut {NoStop}%
\bibitem [{\citenamefont {Kowalevicz~Jr}\ and\ \citenamefont
  {Bucholtz}(2006)}]{KowaleviczJr:2006wc}%
  \BibitemOpen
  \bibfield  {author} {\bibinfo {author} {\bibfnamefont {A.~M.}\ \bibnamefont
  {Kowalevicz~Jr}}\ and\ \bibinfo {author} {\bibfnamefont {F.}~\bibnamefont
  {Bucholtz}},\ }\href {http://www.dtic.mil/docs/citations/ADA456331} {\enquote
  {\bibinfo {title} {{Beam Divergence from an SMF-28 Optical Fiber}},}\
  }\bibinfo {type} {Tech. Rep.}\ (\bibinfo {year} {2006})\BibitemShut {NoStop}%
\bibitem [{\citenamefont {Maeno}\ \emph {et~al.}(2017)\citenamefont {Maeno},
  \citenamefont {Takahashi}, \citenamefont {Nakamura}, \citenamefont {Asano},\
  and\ \citenamefont {Noda}}]{Maeno:2017dv}%
  \BibitemOpen
  \bibfield  {author} {\bibinfo {author} {\bibfnamefont {K.}~\bibnamefont
  {Maeno}}, \bibinfo {author} {\bibfnamefont {Y.}~\bibnamefont {Takahashi}},
  \bibinfo {author} {\bibfnamefont {T.}~\bibnamefont {Nakamura}}, \bibinfo
  {author} {\bibfnamefont {T.}~\bibnamefont {Asano}}, \ and\ \bibinfo {author}
  {\bibfnamefont {S.}~\bibnamefont {Noda}},\ }\href {\doibase
  10.1364/OE.25.000367} {\bibfield  {journal} {\bibinfo  {journal} {Optics
  Express}\ }\textbf {\bibinfo {volume} {25}},\ \bibinfo {pages} {367}
  (\bibinfo {year} {2017})}\BibitemShut {NoStop}%
\bibitem [{\citenamefont {Komma}(2016)}]{Komma:2016vb}%
  \BibitemOpen
  \bibfield  {author} {\bibinfo {author} {\bibfnamefont {J.~K.}\ \bibnamefont
  {Komma}},\ }\emph {\bibinfo {title} {{Optische Eigenschaften von
  Substratmaterialien f{\"u}r zuk{\"u}nftige kryogene
  Gravitationswellendetektoren}}},\ \href
  {https://www.db-thueringen.de/receive/dbt_mods_00029355} {Ph.D. thesis},\
  \bibinfo  {school} {Friedrich-Schiller-Universit{\"a}t Jena} (\bibinfo {year}
  {2016})\BibitemShut {NoStop}%
\bibitem [{\citenamefont {Ross}(2017)}]{Ross:2017tv}%
  \BibitemOpen
  \bibfield  {author} {\bibinfo {author} {\bibfnamefont {M.~P.}\ \bibnamefont
  {Ross}},\ }\emph {\bibinfo {title} {{Bound exciton-assisted spin-to-charge
  conversion of donors in silicon}}},\ \href
  {http://discovery.ucl.ac.uk/1566745/} {Ph.D. thesis},\ \bibinfo  {school}
  {University College London} (\bibinfo {year} {2017})\BibitemShut {NoStop}%
\end{thebibliography}%

\end{document}